# SMART HANDOVER BASED ON FUZZY LOGIC TREND IN IEEE802.11 MOBILE IPV6 NETWORKS

Joanne Mun-Yee Lim[1]  Chee-Onn Chow[2]

[1]Department of Engineering, UCTI (APIIT), Malaysia
jlmy555@gmail.com

[2]Department of Electrical Engineering, University Malaya, Malaysia
cochow@um.edu.my

## ABSTRACT

*A properly designed handoff algorithm is essential in reducing the connection quality deterioration when a mobile node moves across the cell boundaries. Therefore, to improve communication quality, we identified three goals in our paper. The first goal is to minimize unnecessary handovers and increase communication quality by reducing misrepresentations of RSSI readings due to multipath and shadow effect with the use of additional parameters. The second goal is to control the handover decisions depending on the users' mobility by utilizing location factors as one of the input parameters in a fuzzy logic handover algorithm. The third goal is to minimize false handover alarms caused by sudden fluctuations of parameters by monitoring the trend of fuzzy logic outputs for a period of time before making handover decision. In this paper, we use RSSI, speed and distance as the input decision criteria of a handover trigger algorithm by means of fuzzy logic. The fuzzy logic output trend is monitored for a period of time before handover is triggered. Finally, through simulations, we show the effectiveness of the proposed handover algorithm in achieving better communication quality.*

## 1. INTRODUCTION

The basic reason to handover is to prevent communication quality deterioration or disconnection of services by connecting to the Internet at all times. Communication quality degradation can be minimized by choosing the right moment to initiate handover. One of the major challenges in triggering handover at the right moment is choosing the reliable parameters for decision making. There are three main reasons that contribute to the communication quality degradation during handover due to increase false handover trigger alarms. First, due to multipath and shadow effects, misrepresentations of RSSI readings are more common on the communication network quality. Second, the random movement of a mobile node contributes to the misrepresentations of parameters' readings. Third, sudden change of parameters' readings might cause frequent false handover alarms. Therefore, the parameters chosen and the handover algorithm used are the key aspects in the development of solutions supporting mobility scenarios. The parameters chosen should be able to predict communication degradation precisely and trigger timely and reliable handovers by monitoring signal strength and location factors in order to ensure communication quality is either maintained or improved.

Three objectives are identified in this paper in order to achieve improved communication quality during handover. The first objective is to reduce misrepresentations of RSSI readings due to multipath and shadow effect by using additional input parameters. The second objective is to control the handover decisions depending on the users' mobility by employing location factors in handover decision making process. The third objective is to minimize false handover alarms cause by sudden fluctuations of parameters' readings by monitoring the output trend of fuzzy logic for a period of time before making handover decision. With RSSI, speed and distance as the input parameters, a fuzzy logic based handover decision algorithm is developed. This is accomplished by applying fuzzy logic with to evaluate the criteria simultaneously and to initiate handover processes effectively. In our work, by means of simulation, we compare the proposed handover algorithm with some of the existing methods. There are two main contributions in this paper. First, we developed a fuzzy logic based handover algorithm which incorporates RSSI, speed and distance as the input parameters to trigger handovers efficiently and to achieve improved communication quality performance. Second, the outputs of fuzzy logic are monitored for a period of time to observe its pattern of changes so that false handover trigger alarms can be avoided. This paper is organized as follows. Section 2 presents the related work. Section 3 describes the proposed handover scheme. Section 4 provides the details of simulation studies. Section 5 analyzes the results and presents a comparative study with the existing triggering schemes. Section 6 concludes this paper.

## 2. RELATED WORK

Many studies have investigated various numbers of ways to improve handover performance. In this section, we describe the existing handover trigger schemes.

The algorithm developed in [1] was based on received signal strength indicator (RSSI) measurements to predict the next state of mobile node. A new prediction technique called RSSI Gradient Predictor was used to detect the change of state in the RSSI values. The predictor was proved to be efficient for applications of real time video in a wireless network that combined wireless fidelity (WiFi) and worldwide interoperability for microwave access (WiMAX) technologies.

In [2], a handoff ordering method based on packet success rate (PSR) for multimedia communications in wireless networks was proposed. A prediction was done on the remaining time for each session to reach its minimum PSR requirement. Results showed that PSR could effectively improved the handoff call dropping probability with little increased of the new call blocking probability.

In [3], the different aspects of handoffs designs and performance related issues were discussed. A vertical handoff decision function (VHDF) that provided handoff decisions while roaming across heterogeneous wireless network was implemented. VHDF utilized cost of service, security, power consumption, network conditions and network performance as the parameters in making handover decisions. VHDF managed to increase the throughput especially in situations where the background traffic varied.

A handoff algorithm which triggered handovers based on both, distance from a mobile station to the neighbouring base stations and relative signal strength (RSS) measurement was proposed [4]. The algorithm performed handoff when the measured distance exceeded a given threshold and when the RSS exceeded a given hysteresis level. However, the performance of the proposed algorithm was less efficient under worst case conditions in comparison to other algorithms. This could be resolved by employing a high accuracy location method such as differential global positioning system (GPS) or real time kinematic global positioning system (GPS).

A fuzzy normalization concept in handover decisions within a heterogeneous wireless network was introduced [5]. The fuzzy inputs, relative signal strength (RSS), velocity and system loading were used as the input parameters for the proposed fuzzy normalization (FUN) – handover decision strategy (HODS), FUN-HODS. Finally, FUN-HODS proved to be able to balance the system loading while reducing handover failures which were usually caused by mobile node's velocity and weak RSS.

In [6], the authors stated that the performance degradation in mobile nodes (MNs) were usually due to reduction of signal strength caused by mobile nodes' movement, intervening objects and radio interference with other wireless local area networks (WLANs). By employing file transfer protocol (FTP) and voice over internet protocol (VoIP) applications, the usage of signal strength and number of frame retransmission as the handover triggers were investigated through experiments in a real environment. The results showed that signal strength could not promptly and reliably detect the degradation of communication quality in both FTP and VoIP communications when the signal strength was affected by MN's movement or intervening objects. However, the number of frame retransmissions was capable of detecting the degradation of communication quality of a wireless link due to MN's movement and intervening objects. Therefore, it was suggested that the number of frame retransmissions, unlike signal strength, was able to detect the communication quality degradation caused by radio interference and reduction of signal strength.

## 3. FUZZY LOGIC TREND (RSSI, SPEED AND DISTANCE), FL TREND (RSD)

In this section, we describe the proposed handover trigger scheme, Fuzzy Logic Trend (RSSI, Speed and Distance), FL Trend (RSD). The list of precise steps and the order of computations in the FL Trend (RSD) is shown in figure 1. The FL Trend (RSD) starts at iteration of i=0 every time the mobile node attaches itself to a new access point. Subsequently, FL Trend (RSD) starts obtaining RSSI, speed and distance readings at every interval of B. The RSSI, speed and distance collected are used as the input parameters in the FL Trend (RSD) as the representation of the current communication quality. Next, FL Trend (RSD) used these input readings to collect respected input scores based on fuzzy logic input membership functions as shown in figure 2(a) – (c). The range of distance, speed and RSSI shown in figure 2 (a) – (c) are based on a human walking or running scenario in a campus environment. Based on fuzzy logic rule matrix shown in table 1, the respected output scores are obtained for each rule matrix. These output scores are used to obtain the final output score which is used to decide whether to handover or not to handover. The final output score is calculated based on equation (1):

$$Score = \frac{(-100)*Not\ handover\ score + (100)*Handover\ score}{Not\ handover\ score + Handover\ score}$$

(1)

where not handover score is calculated from equation (2):

$$Not\ handover\ score = \sqrt{NH1^2 + NH2^2 + \cdots \ldots \ldots \ldots + NH13^2}$$

(2)

and handover score is calculated from equation (3):

$$Handover\ score = \sqrt{H1^2 + H2^2 + \cdots \ldots \ldots \ldots + H14^2}$$

(3)

A sliding window is set to monitor the FL Trend (RSD)'s output scores trend as shown in figure 3. If within N intervals, a threshold of $TH_{HO}$ has not been reached, the algorithm adds to the iteration i and continues to monitor the input parameters, RSSI, speed and distance. Otherwise, the handover process is initiated. After the handover trigger has been initiated, next, a delay of W, calculated in equation (4) where speed is the speed of the moving MN and distance, D is the distance that the MN has travelled is introduced. This delay prevents mobile node from triggering multiple handovers after the node attaches to a new access point with the intention that ping pong effect could be prevented.

$$Delay, W = \frac{Distance, D}{Speed} \quad (4)$$

If the speed is low, the delay is longer as it foresees that the mobile node takes longer time to move to the edge of the coverage. However, if the speed is high, the delay is shorter because the probability of the mobile node moving to the edge in a short period of time is higher. Thus, a shorter delay ensures that the parameters are closely monitored. After this delay period, the value i is set back to zero as the mobile node has now been connected to a new access point. The speed and distance in the equation can be obtained using methods proposed by [7] that uses the difference between two consecutive signal updates and free space path loss (FSPL) equations to estimate the speed of a node. Using FSPL equation, the covered distance can easily be calculated if the attenuation of signal is known. Using the time difference between two mobile nodes at two different positions, speed is calculated. In another method, information about the angle of arrival of the signal is used to calculate the estimated speed using cosine theorem. The estimation of speed can be calculated at both nodes; the mobile node or the network node. The travelling period can be obtained using the time difference between two different nodes at two different positions. With this information, the speed and distance of the node can be calculated.

Table 1
FL Trend (RSD) Rule Matrix

|  | Input of Fuzzy Logic | | | |
| --- | --- | --- | --- | --- |
| Rule | RSSI | Speed | Distance | Output |
| 1 | High | High | High | Not handover, NH1 |
| 2 | High | Medium | High | Not handover, NH2 |
| 3 | High | Low | High | Not handover, NH3 |
| 4 | Medium | High | High | Handover, H1 |
| 5 | Medium | Medium | High | Handover, H2 |
| 6 | Medium | Low | High | Handover, H3 |
| 7 | Low | High | High | Handover, H4 |
| 8 | Low | Medium | High | Handover, H5 |
| 9 | Low | Low | High | Handover, H6 |
| 10 | High | High | Medium | Not handover, NH4 |
| 11 | High | Medium | Medium | Not handover, NH5 |
| 12 | High | Low | Medium | Not handover, NH6 |
| 13 | Medium | High | Medium | Handover, H7 |
| 14 | Medium | Medium | Medium | Handover, H8 |
| 15 | Medium | Low | Medium | Not handover, NH7 |
| 16 | Low | High | Medium | Handover, H9 |
| 17 | Low | Medium | Medium | Handover, H10 |
| 18 | Low | Low | Medium | Handover, H11 |
| 19 | High | High | Low | Not Handover, NH8 |
| 20 | High | Medium | Low | Not handover, NH9 |
| 21 | High | Low | Low | Not handover, NH10 |
| 22 | Medium | High | Low | Not Handover, NH11 |
| 23 | Medium | Medium | Low | Not Handover, NH12 |
| 24 | Medium | Low | Low | Not handover, NH13 |
| 25 | Low | High | Low | Handover, H12 |
| 26 | Low | Medium | Low | Handover, H13 |
| 27 | Low | Low | Low | Handover, H14 |

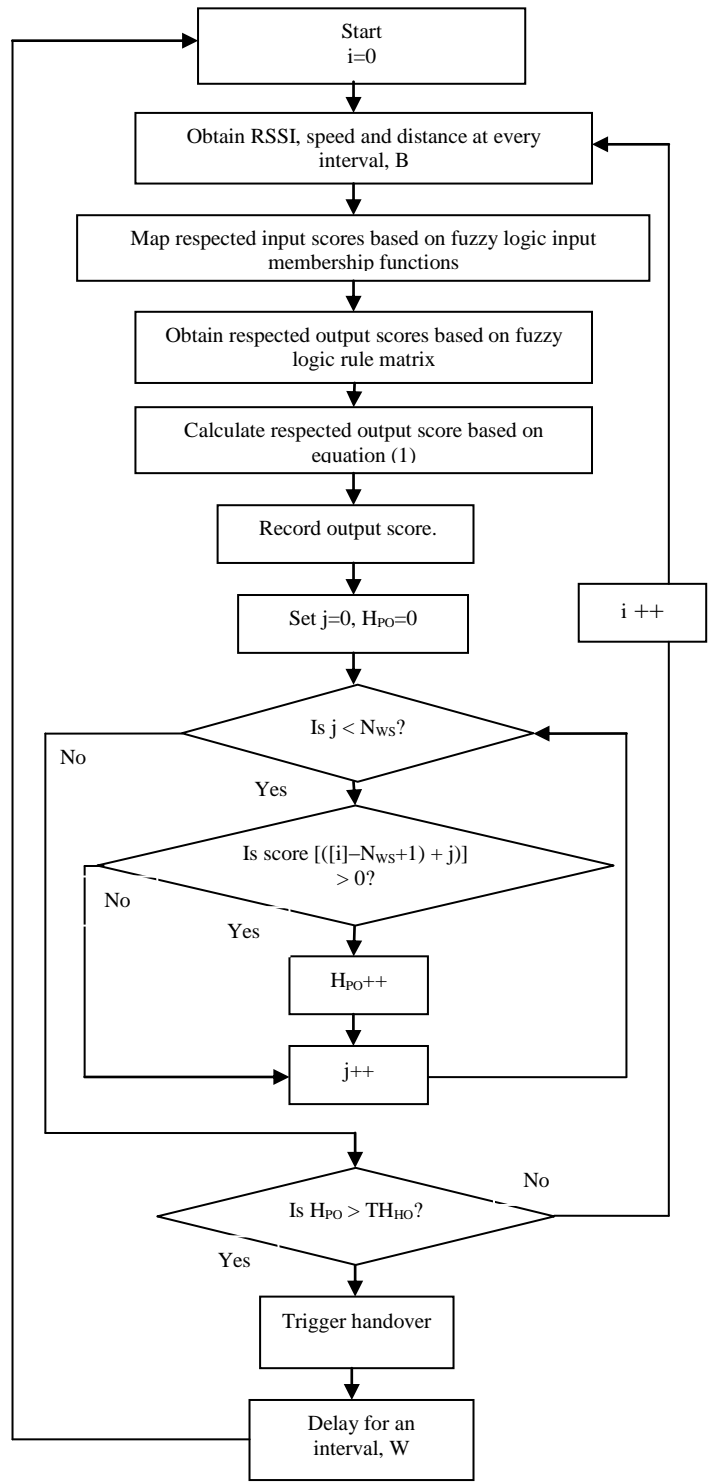

Figure 1: FL Trend (RSD) Flow Chart

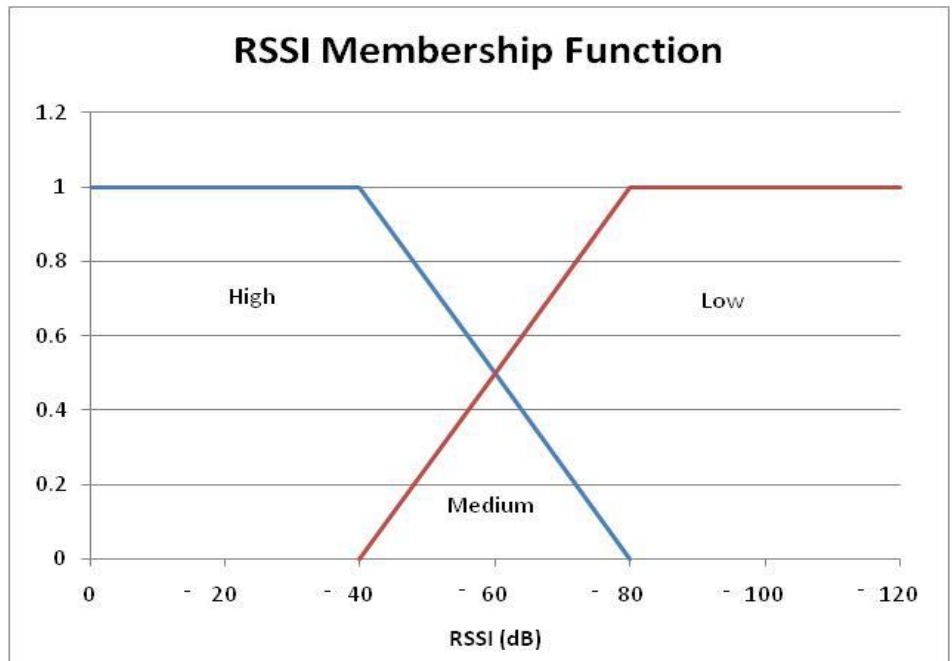

Figure 2(a): RSSI Membership Function

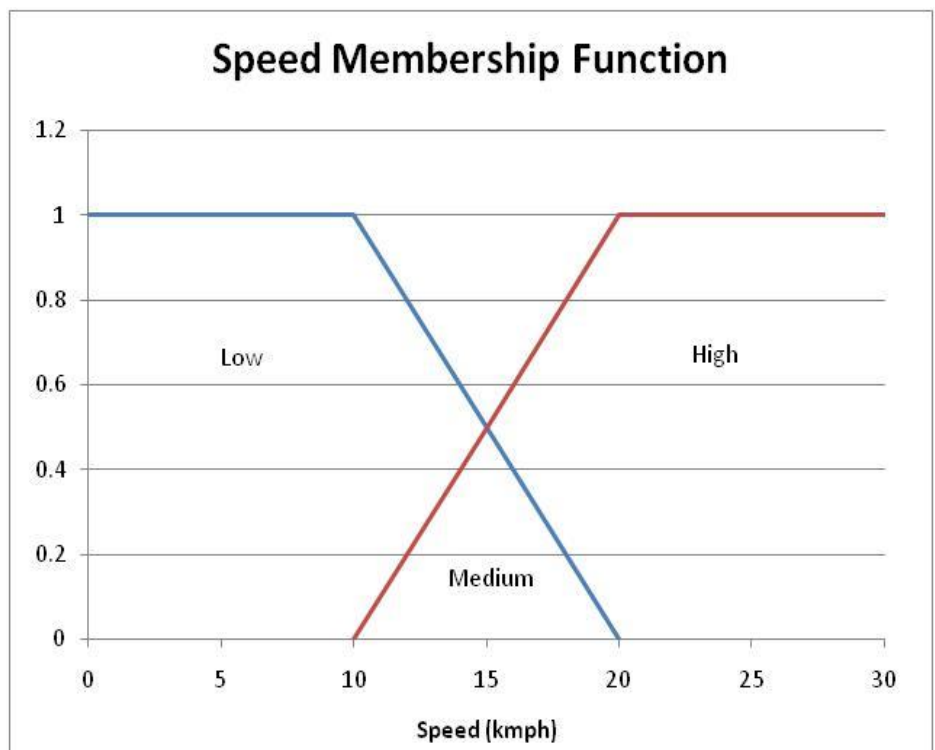

Figure 2(b): Speed Membership Function

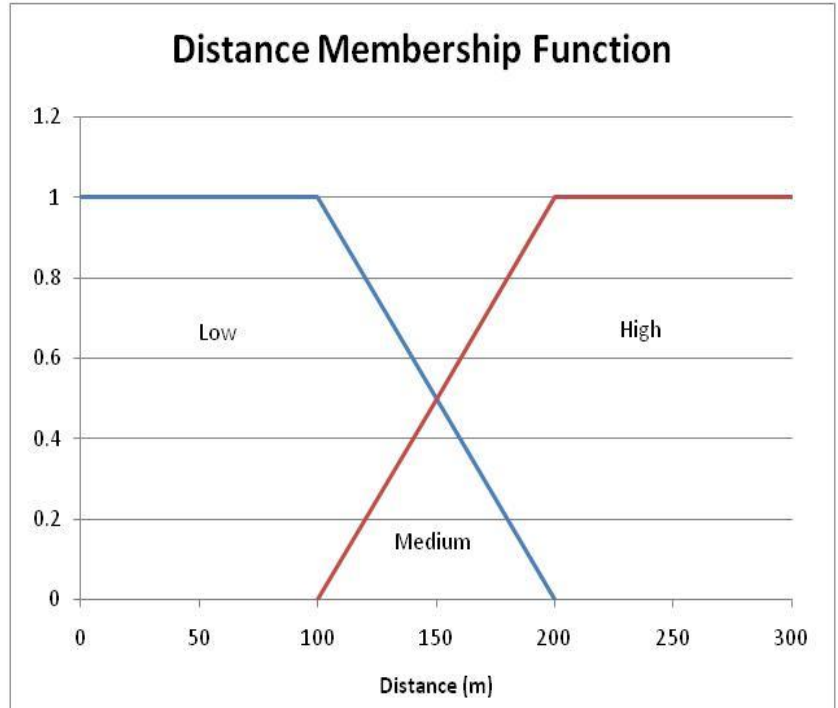

Figure 2(c): Distance Membership Function

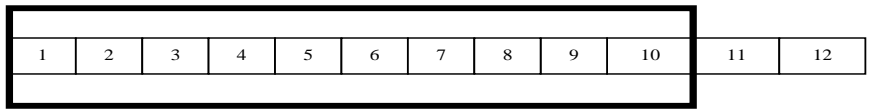

First interval, where ten consecutive [i] scores are being monitored.

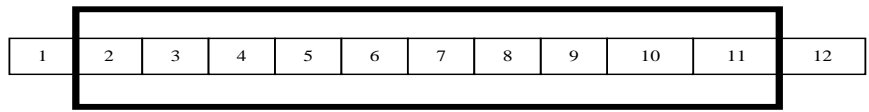

Second interval, where another ten consecutive [i] scores are being monitored

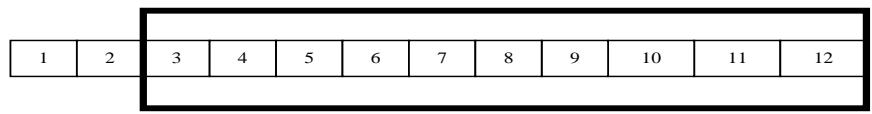

Third interval, where another ten consecutive [i] scores are being monitored

Figure 3: Timing Chart of a Sliding Window where $N_{WS}=10$.

# 4. SIMULATION STUDIES

In order to evaluate the performance of FL Trend (RSD), we designed and implemented the following scenario as shown in figure 4. Mobile node (MN) traversed randomly in the assigned space within a time limit. The simulation was run at different speeds ranging from 5kmph to 30kmph. The low speed range (5kmph to 15kmph) defined in this simulation is to replicate the human walking scenario whereas the high speed range (16kmph to 30kmph) defined in this simulation is to replicate a human running or a vehicle driving in an urban area scenario. The access points were intentionally placed overlapping one another to generate interference. The reason behind this implementation was to create a realistic simulation scenario and also to test the ability of different parameter to detect communication degradation. This simulation was run on Omnet 4.0 simulator [8]. Mobile node started from the home agent and traversed randomly across the network. Once a handover triggering signal was received, mobile node performed handshake process with the selected access point. CN received updates from MN. MN then started sending traffic flow to the new interface where MN was connected. Subsequently, MN began sending data transmission to the corresponding node (CN) at a constant bit rate with the packet size of 1000 bytes at the interval of F, 0.5s whereas CN began sending data transmission at a constant bit rate with the packet size of 1000bytes at the interval of C, 0.08s. When MN detected the necessity to handoff once more, the handover algorithm was triggered and MN continued the handover process to ensure the continuity of communication session. The simulation was run on ten different scenarios and the averaged results were recorded. The results obtained were tabulated and averaged. The parameters employed in this simulation were shown in table 2. In this comparative study, we evaluated four handover trigger algorithms; RSSI Threshold, Change of RSSI, FR Threshold and FL Trend (RSD). RSSI Threshold scheme's threshold was set at 75dB [9]. An analytical model derived in [10] was used to estimate the threshold for RSSI Threshold scheme link going trigger. Based on Fritz path loss model in equation (5):

$$\left[\frac{Prx(d)}{Prx(do)}\right] dB = -10\beta \log\left[\frac{d}{do}\right] \quad (5)$$

where $Prx(d)$ was the received signal power level in Watts, $Prx(do)$ was the received power at the close-in reference distance, $d$ was the distance between the transmitter and receiver, $do$ was determined using the free space path loss model and $\beta$ was the path loss exponent, the threshold was calculated. With this, a threshold of 86dB was obtained. Hence, according to [9, 10], a threshold of 75dB was used as the threshold limit for RSSI Threshold algorithm.

Table 2: Simulation Parameters

| Parameter | Value |
|---|---|
| Transmitter Power | 50mW |
| Wavelength (λ) | 0.125m |
| Path Loss Exponent | 2 |
| Radio Carrier Frequency | 2.4GHz |
| Minimum Channel Time | 1s |
| Maximum Channel Time | 3s |
| CN Data Interval, C | 0.08s |
| MN Data Interval, F | 0.5s |
| Distance, D | 100m |
| Update Interval, B | 0.1s |
| Sliding Window Size, $N_{WS}$ | 10 |
| Threshold, $TH_{HO}$ | 7 |

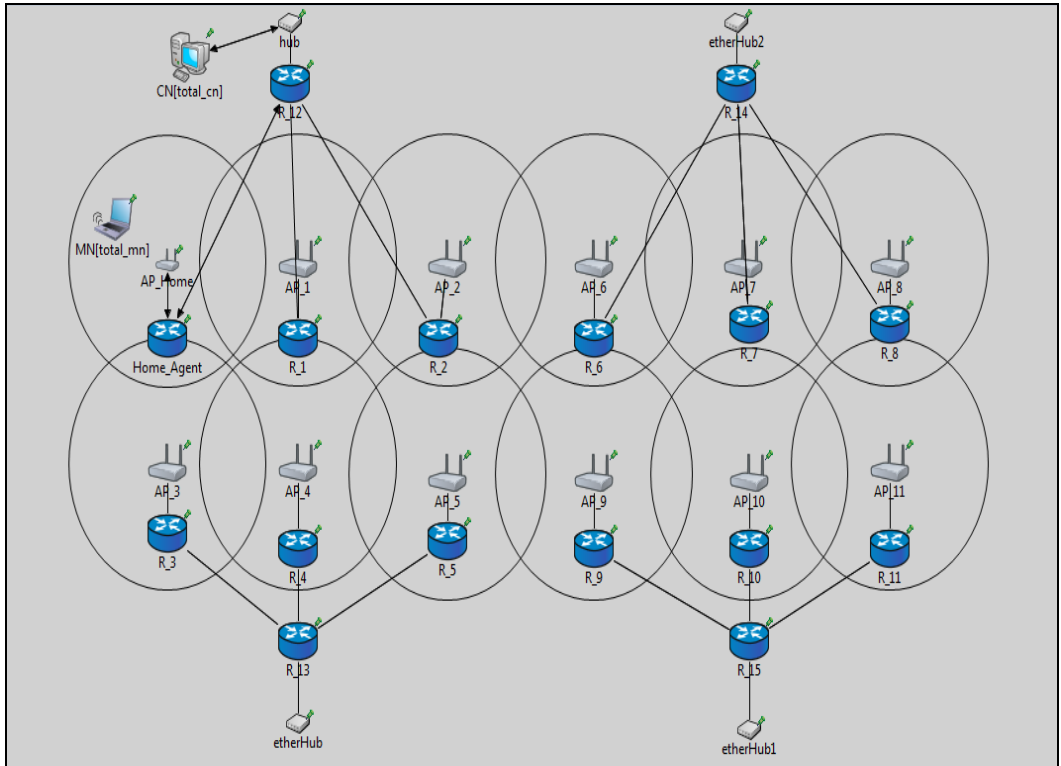

Figure 4: Simulation Scenario

## 5. RESULTS AND DISCUSSION

In this section, we evaluate four handover trigger algorithms. Through performance analysis we analyze the existing handover trigger algorithms and compare them with our proposed handover trigger algorithm. To achieve seamlessness in wireless networks, the algorithm has to satisfy multiple objectives. We advocate that efficient handover trigger algorithms should have the following characteristics to support seamless handovers in a wireless network. These characteristics include being able to guarantee low packet loss, high throughput and low packet delay during handover. All these should be achieved with minimal number of handovers in order to attain improved communication quality. We study the performance of the algorithms with data traffic. The evaluations of different schemes cover the following dimensions: (i) Number of handovers. (ii) Average packet loss. (iii) Average throughput. (iv) Minimum packet delay. (v) Maximum packet delay. (vi) Mean packet delay. Number of handovers depicts the process of transferring an ongoing data session; therefore it is crucial to be evaluated as increased of number of handovers affects the communication quality experienced by the users. Packet loss causes the transmitted packets to fail to arrive at their designated destination which in turn result in communication errors. Average throughput defines the average success rate of packets delivery over a period of time, thus high throughput is desirable to achieve high data rates. Minimum packet delay is the minimum latency of each link where low minimum packet delay is preferred as it indicates shorter time required by a packet to reach its designated destination. Maximum packet delay represents the maximum latency of each link where high maximum packet delay is not favored as it indicates that the communication link is probably experiencing congestion or abrupt change of communication quality. Mean packet delay defines the average latency of packets where high mean packet delay indicates bad communication quality which requires MN to perform a handover in order to maintain the ongoing communication session.

It is pointed out that RSSI threshold is unable to guarantee improved communication quality. This is shown in our comparative study. RSSI threshold attains high number of handovers as shown in figure 5, high packet loss as shown in figure 6 and 7, low throughput as shown in figure 8 and 9 and high packet delay as shown from figure 10 to 15. Due to multipath and shadow fading, RSSI sometimes fluctuates abruptly even when the established connection between the mobile node and access point is still adequate. Thus, this forces unnecessary handovers to take place which decreases the communication performance. RSSI Threshold only triggers handovers when the RSSI threshold is reached. Instead, RSSI Threshold does not monitor the RSSI readings throughout a period of time. Therefore, a sudden change in the RSSI readings would have caused handovers to be triggered when in actual scenario, it is not necessary as the RSSI readings are just experiencing a sudden drastic drop for a short period of time. Hence, solely based on RSSI Threshold to trigger handovers has the potential to cause unnecessary handovers which degrades the communication quality performance.

Through evaluations, we observe that the Change of RSSI achieves low packet loss as shown in figure 6 and 7 and high throughput as shown in figure 8 and 9 at the cost of high packet delay as shown in figure 10 to 15 and high number of handovers as shown in figure 5. This is especially true in the low speed range (5kmph to 15kmph). The conventional methods use RSSI as the handover trigger parameter. This is because RSSI triggers handover when the signal strength becomes weak. However, at times, the RSSI gives wrong representation of the current communication link quality as RSSI is easily affected by multipath and shadow fading. Therefore, the change of RSSI is being monitored through a period of time before the decision to handover is made in the Change of RSSI algorithm. If the change of RSSI indicates the necessity to trigger handover, handover is carried out. However, the drawback of using change of RSSI is that unnecessary handover might be triggered due to the incapability of RSSI to detect radio interference which causes degradation of communication quality. Therefore, we observe high number of handovers in Change of RSSI at all range of speed which results in the increment of overall packet delay. However, more handovers results in higher probability of mobile node being handover to close proximity access points. Thus, this improves the throughput and packet loss slightly within the low speed range as less packet delay is incurred due to nearer access point connectivity. At high speed range (16kmph to 30kmph), mobile node moves around at higher speeds which results in abrupt change of RSSI in a short period of time. This results in increase number of handovers. Subsequently, this in turns incurs higher packet delay as the distance between the connected access point and mobile node changes abruptly due to the increase of speed. The increase probability of unnecessary handovers causes handovers to take place to the less efficient access points. Due to the abrupt change in the mobility of mobile node caused by the increase of speed, the probability of wrongly chosen access point as a result of false handover trigger alarms is higher. Therefore, the false handover trigger alarms causes handovers to take place to the inefficient and distant access points. This in turn causes high packet loss and low throughput at the high speed region.

It was shown in [6] that FR Threshold has the ability to detect radio interference much better as compared to RSSI readings. As pointed out in the low speed range, FR Threshold shows 5% to 20% lower number of handovers as shown in figure 5, lower packet loss as shown in figure 6 and 7 and higher throughput as shown in figure 8 and 9 in comparison with FL Trend (RSD) at the cost of higher packet delay as shown from figure 10 to 15. However, at the high speed region, FL Trend (RSD) outperforms FR Threshold. This is because FR Threshold does not monitor the frame retransmission readings throughout a period of time. Therefore, sudden change of frame retransmission would have caused unnecessary handovers to take place which causes degradation of communication quality. At high speed region, mobile node moves drastically in a high speed manner which causes constant abrupt change of frame retransmission readings. This causes frame retransmission threshold to be reached constantly, thus causing unnecessary handovers to be triggered. This in turn causes high packet loss and low throughput which is exceptionally obvious in the high speed region. FL Trend (RSD) outperforms FR Threshold in the high speed region in eliminating unnecessary handovers by taking location and speed of mobile node into consideration as the input parameters. Besides that, FL Trend (RSD) monitors the change of fuzzy logic outputs in a known period of time which eliminates inefficient handover decisions by avoiding unnecessary handovers that are triggered due to sudden change of parameters readings.

According to figure 5, FL Trend (RSD) is able to perform better in guaranteeing lower number of handovers as shown in figure 5; lower packet loss as shown in figure 6 and 7 and higher throughput as shown in figure 8 and 9. This is especially true in the high speed range (16kmph to 30kmph). Our comparative study shows that FL Trend (RSD) is able to attain lower packet delay as seen from figure 10 to 15. FL Trend (RSD) considers RSSI, speed and distance as the inputs to obtain accurate output decisions of whether handover is necessary. RSSI fluctuates abruptly due to multipath and shadow fading. However, RSSI readings are less affected by radio interference. As such, RSSI readings might not be dependable in situations where radio interference is strong. However, in such situation where radio interference and multipath fading exist, with distance and speed as the input parameters, the location of mobile node can be identified. With this information, any misinterpretations of RSSI can be minimized. Therefore, with the use of these three parameters, RSSI, speed and distance, unnecessary handovers are minimized. Hence, FL Trend (RSD) manages to achieve lower number of handovers, lower packet loss, lower packet delay and higher throughput which is especially obvious in the high speed region.

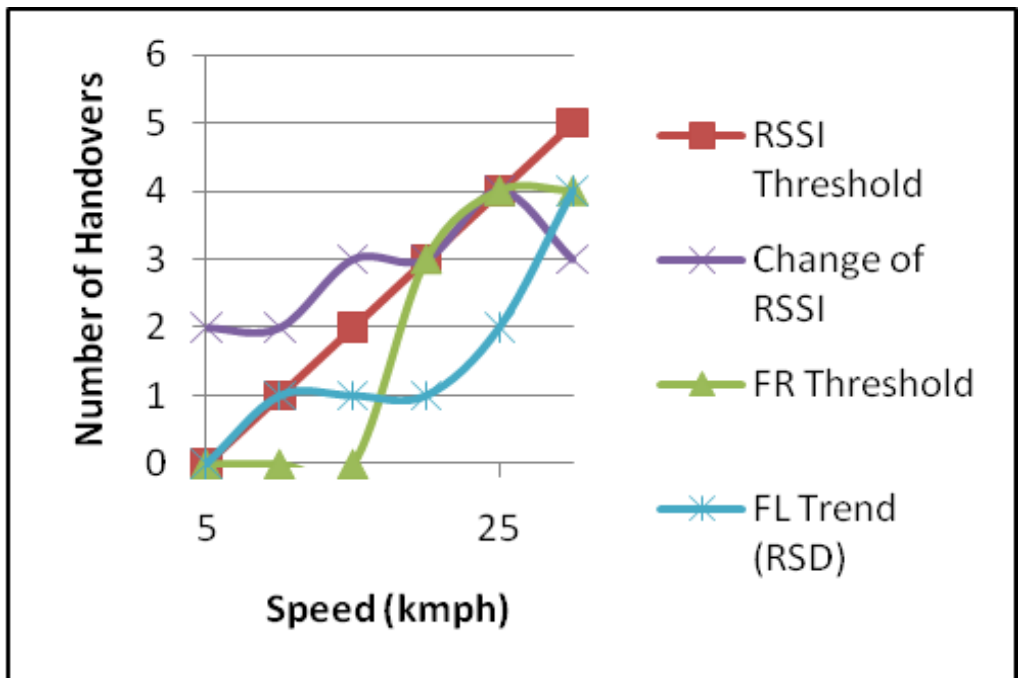

Figure 5: Number of Handovers versus Speed (kmph) at Mobile Node, MN

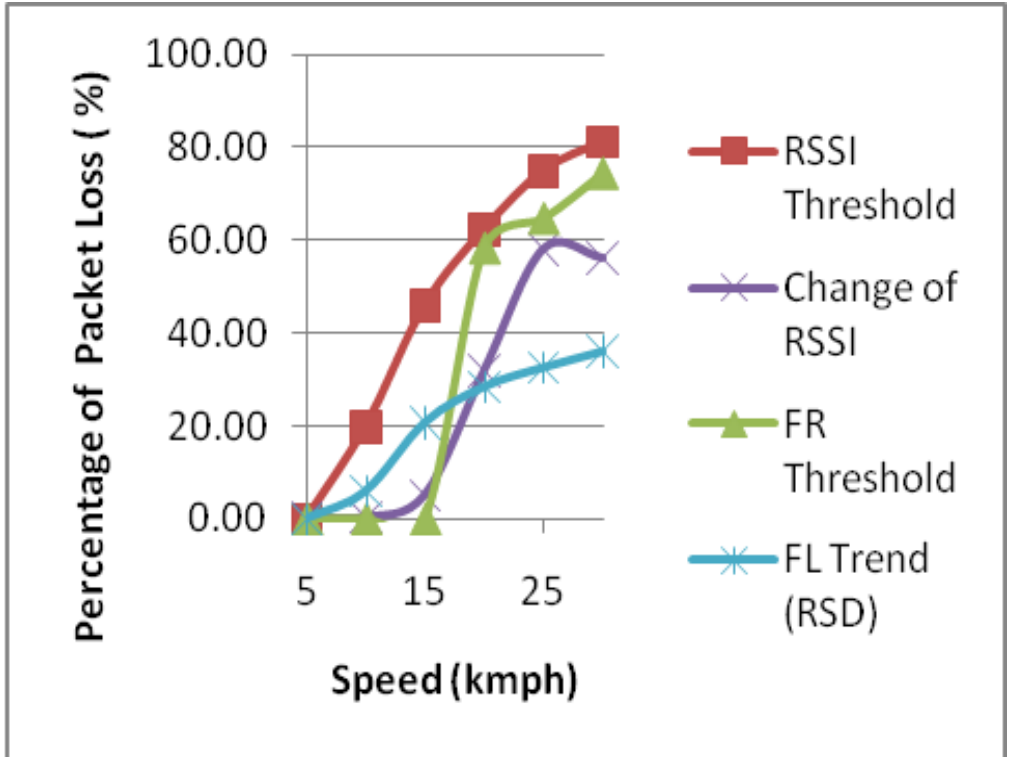

Figure 6: Percentage of Packet Loss (%) versus Speed (kmph) at Mobile Node, MN

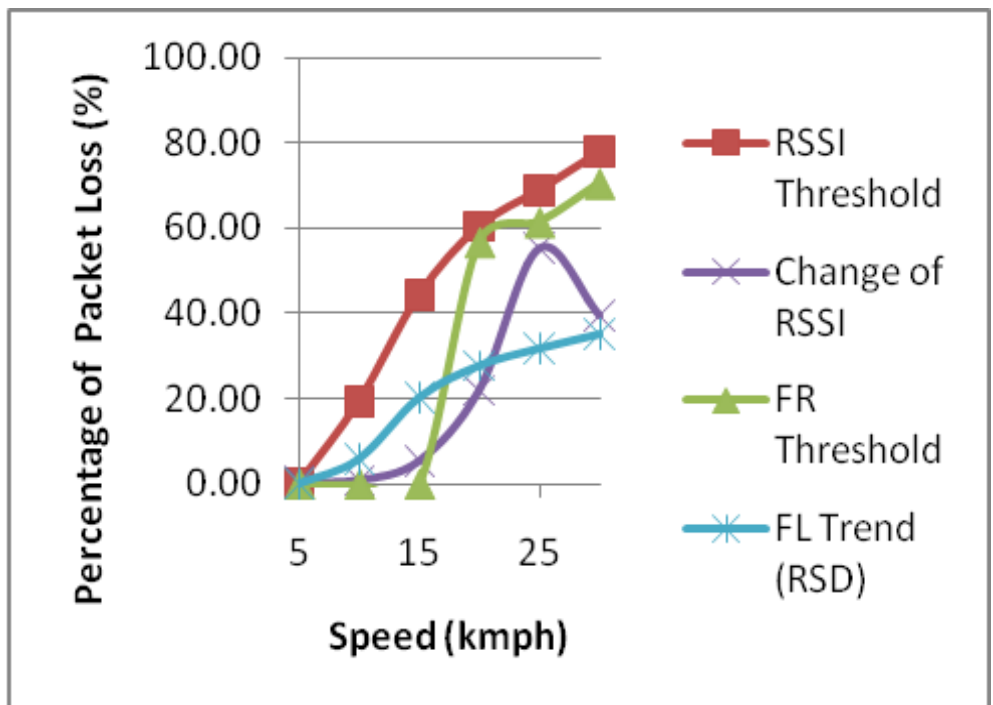

Figure 7: Percentage of Packet Loss (%) versus Speed (kmph) at Corresponding Node, CN

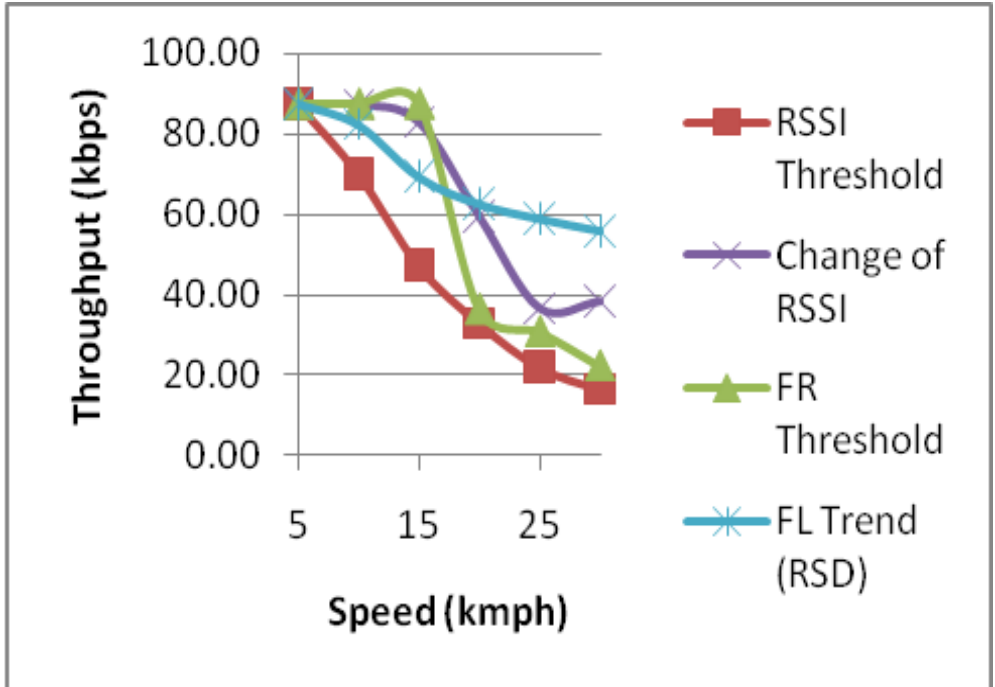

Figure 8: Throughput (kbps) versus Speed (kmph) at Mobile Node, MN

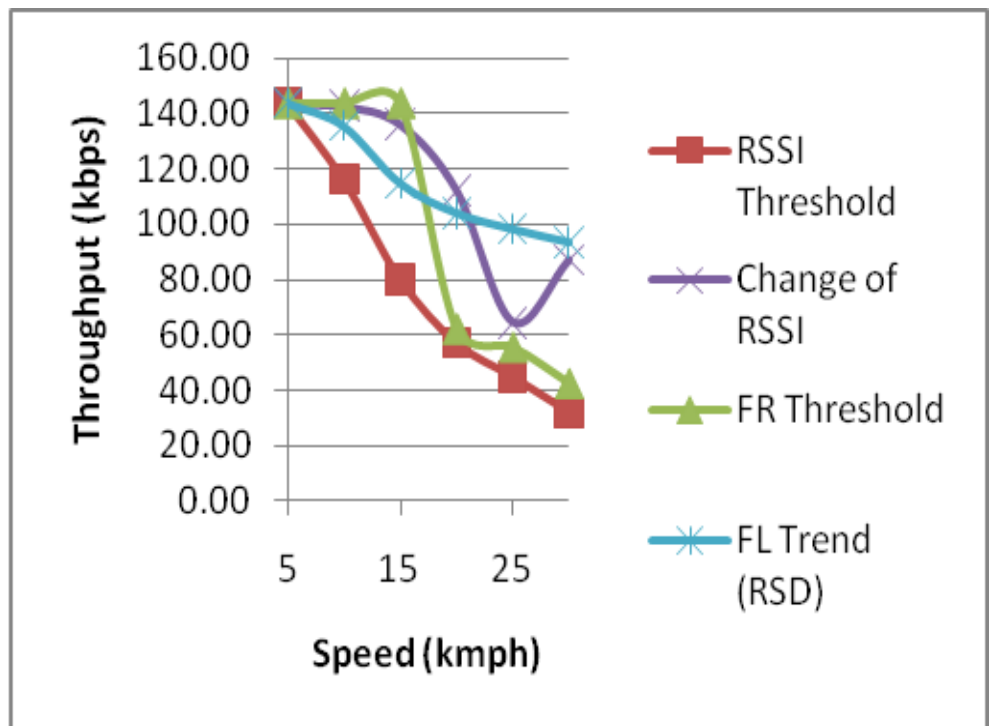

Figure 9: Throughput (kbps) versus Speed (kmph) at Corresponding Node, CN

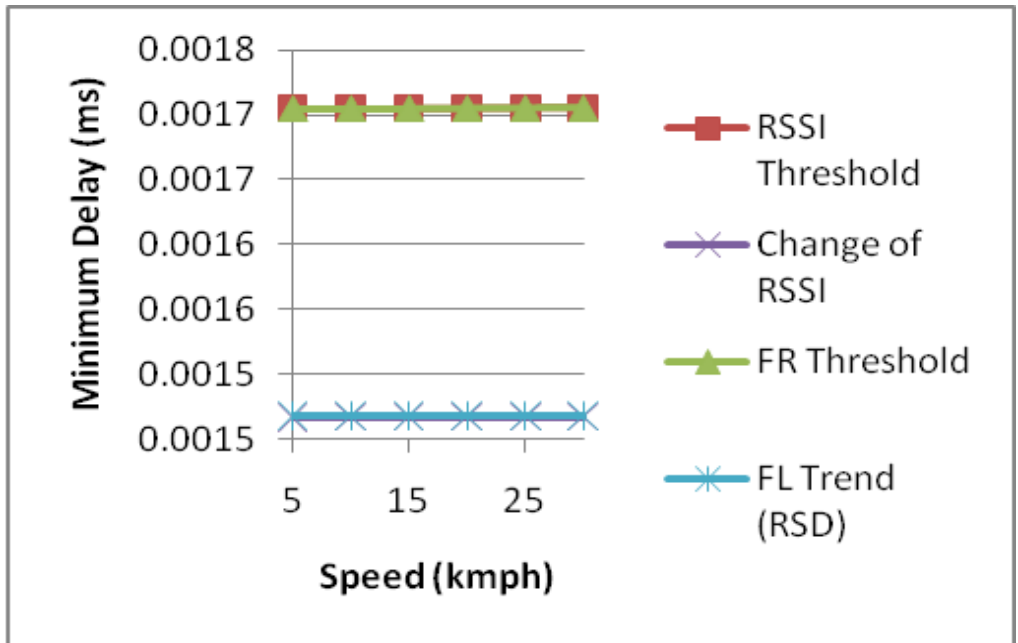

Figure 10: Minimum Delay (ms) versus Speed (kmph) at Mobile Node, MN

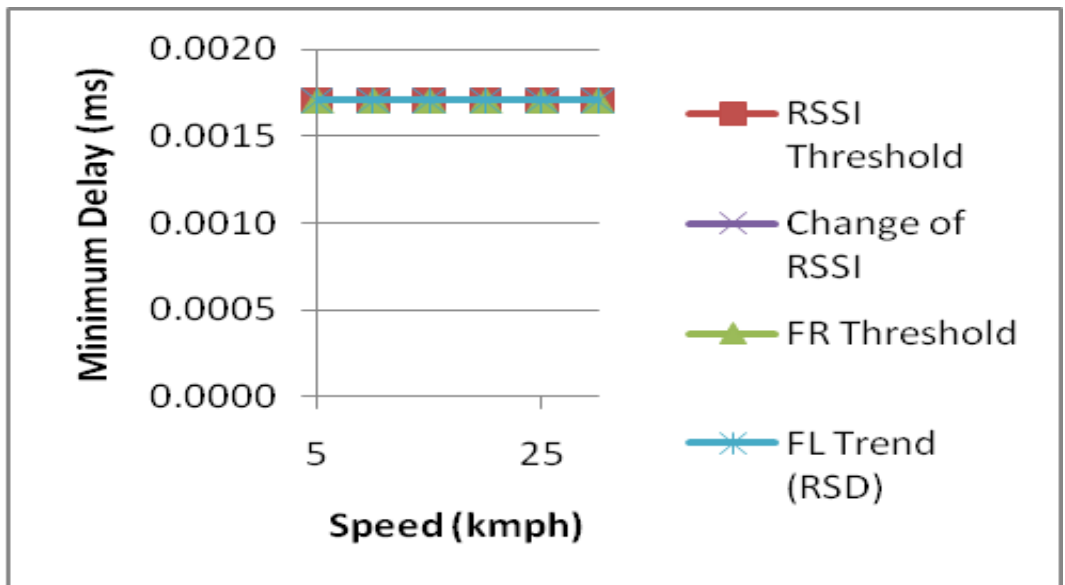

Figure 11: Minimum Delay (ms) versus Speed (kmph) at Corresponding Node, CN

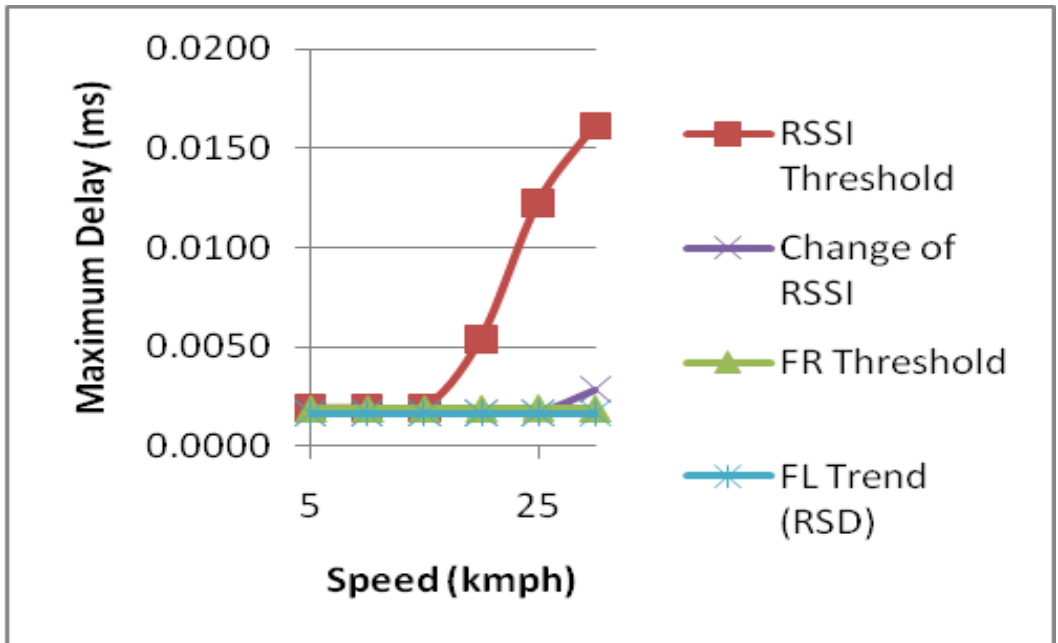

Figure 12: Maximum Delay (ms) versus Speed (kmph) at Mobile Node (MN)

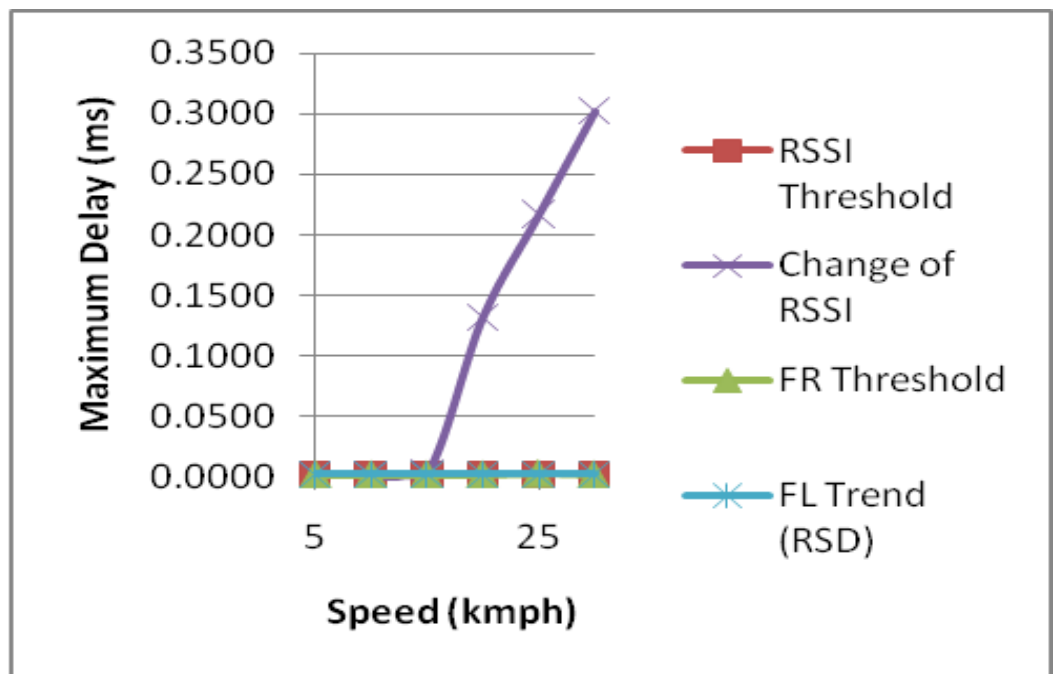

Figure 13: Maximum Delay (ms) versus Speed (kmph) at Corresponding Node, CN

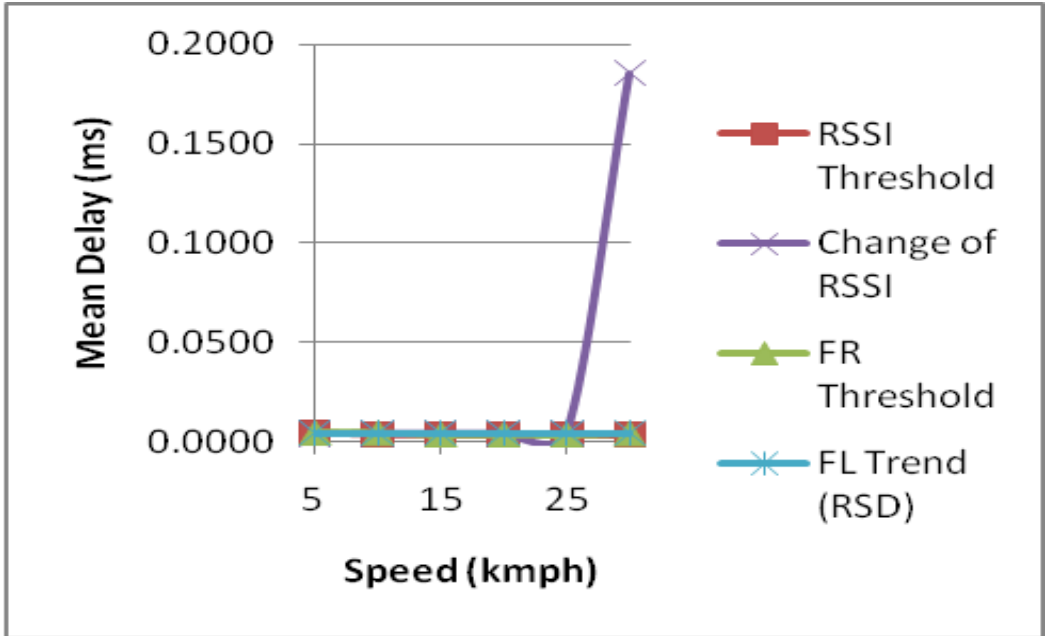

Figure 14: Mean Delay (ms) versus Speed (kmph) at Mobile Node, MN

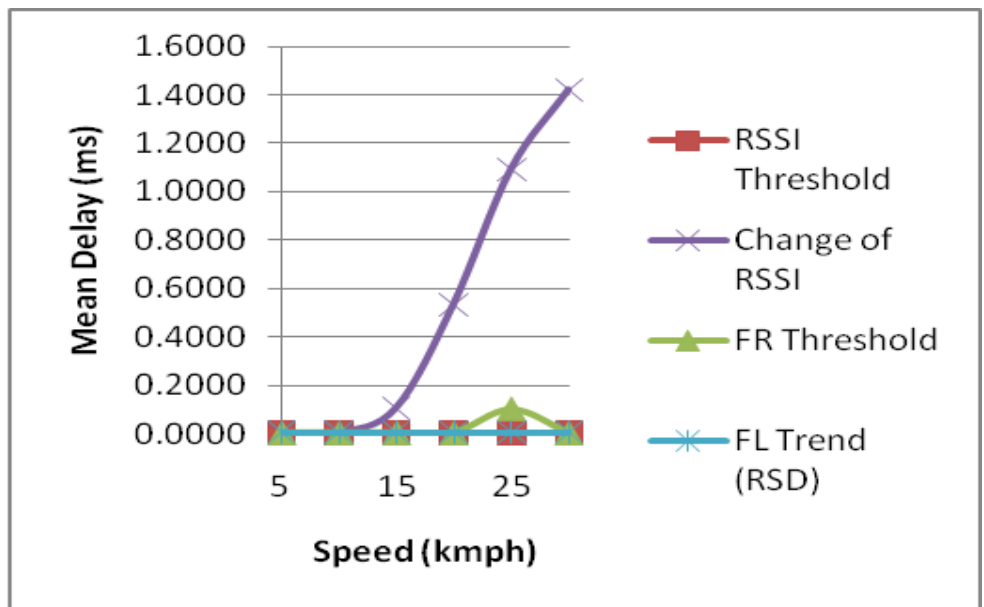

Figure 15: Mean Delay (ms) versus Speed (kmph) at Corresponding Node, CN

## 6. SUMMARY

In this section, we present a comparative performance of four handover trigger algorithms; RSSI Threshold, Change of RSSI, FR Threshold and FL Trend (RSD). Handovers are required to maintain the communication quality of a session. Traditionally it was pointed out that RSSI has the capability of detecting communication quality degradation and triggers handovers. In a wireless network, RSSI changes abruptly due to many factors such as multipath and shadow fading. Therefore, a more challenging problem is to incorporate several parameters to assist RSSI to minimize misrepresentations of the communication quality. In this context, we propose to use RSSI, speed and distance as the decision factors in a fuzzy logic system to optimize performance quality. In addition, we also propose to monitor the output trend of fuzzy logic in FL Trend (RSD) before making handover decisions in order to minimize unnecessary handovers due to sudden fluctuation in the parameters' readings. In comparison with RSSI Threshold, FL Trend (RSD) achieves lower number of handovers, lower packet loss, higher throughput and lower packet delay at all speed ranges. In comparison with Change of RSSI, at low speed (5kmph to 15kmph) FL Trend (RSD) achieves lower number of handovers and lower handover delay at the cost of 5% to 20% higher packet loss and 5% to 20% lower throughput. However, at the high speed range (16kmph to 30kmph), FL Trend (RSD) outperforms Change of RSSI in terms of number of handovers, packet loss, throughput and packet delay. In comparison with FR Threshold, in the low speed range, FL Trend (RSD) achieves lower packet delay at the cost of 5% to 20% lower number of handovers, higher packet loss and lower throughput. However, at high speed, FL Trend (RSD) outperforms FR Thresholds with lower number of handovers, lower packet loss, higher throughput and lower packet delay. The slight decrease of FL Trend (RSD) performance in the low speed region is generally due to the inefficiency of RSSI to detect radio interference. However, this is not the case in the high speed region because in the high speed region, distance readings contribute more towards decision making in FL Trend (RSD) which helps FL Trend (RSD) to make better and more effective handover triggers to achieve improved communication quality.

## 7. CONCLUSION

In wireless networks, handovers are necessary to maintain the communication quality. Traditionally, handovers occur when the signal strength of the serving access point drops below a certain threshold value. However, in mobile IPv6 environment, users are predicted to move constantly. Therefore, location factors are necessary to be considered when making handover decisions besides signal strength because location factors give an insight of the location of mobile node from the access points. Addressing the drawbacks of sudden change in parameters in a radio propagation environment, the outputs fuzzy logic in FL Trend (RSD) is being monitored for a fixed period of time before handover decisions are made. In this paper, we showed that FL Trend (RSD) outperforms the existing handover trigger algorithms in terms of number of handovers, packet loss, throughput, minimum packet delay, maximum packet delay and mean packet delay. This is especially obvious in the high speed topologies (16kmph to 30kmph). Additionally, in the low speed topologies (5kmph to 15kmph), FL Trend (RSD) outperforms RSSI Threshold in all aspects, achieves lower number of handovers and packet delay at the cost of 5% to 20% increase in packet loss and decrease in throughput as compared to Change of RSSI and achieves lower packet delay at the cost of 5% to 20% increase in the number of handovers, increase in packet loss and decrease in throughput as compared to FR Threshold. In general, FL Trend (RSD) achieves improved communication quality by attaining lower number of handovers, lower packet loss, lower packet delay and higher throughput which is especially obvious in the high speed region. Analysis and simulations show that FL Trend (RSD) is able to create timely and reliable handover triggers to achieve improved communication quality performance which is certainly apparent in the high speed region.

**Authors' Biographies**

1. Joanne Mun-Yee Lim

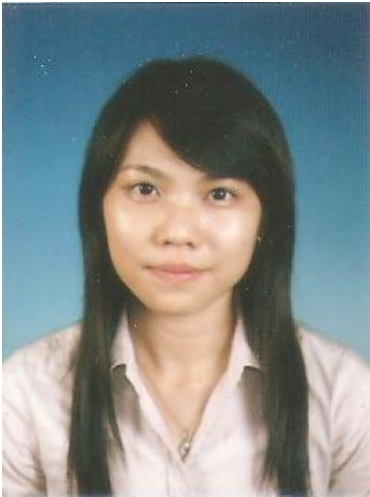

Joanne Mun-Yee Lim received her Bachelor of Engineering (honors) from Monash University in 2008. In 2010, she joined the Department of Electrical Engineering as a tutor in and pursued her Master of Engineering Degree in Electrical Engineering in University Malaya. She is currently a lecturer in UCTI (APIIT). Her research interests include Mobile IPv6 based networks and applications.

2. Chee-Onn Chow

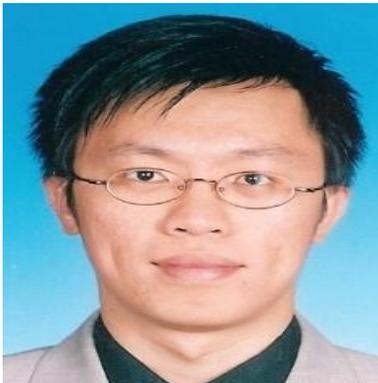

Chee-Onn Chow received his Bachelor of Engineering (honors) and Master of Engineering Science degrees from University of Malaya, Malaysia in 1999 and 2001. He received his Doctorate of Engineering from the Tokai University, Japan in 2008. He joined the Department of Electrical Engineering as tutor in 1999, and subsequently been offered a lecturer position in 2001. He is currently a Senior Lecturer in the same department since 2008. His research interests include multimedia applications and design issues related to next generation networks.